\documentclass[
 reprint,
 superscriptaddress,
 amsmath,amssymb,
 aps,
 longbibliography,
 floatfix
]{revtex4-2}

\usepackage{graphicx}
\usepackage{pgf}
\usepackage{pgfplots}
\usepackage{tikz}
\usepackage{dcolumn}%
\usepackage{bm}%
\usepackage{siunitx}%
\usepackage{hyperref}
\usepackage{import}
\usepackage{mathtools}
\usepackage{subcaption}%
\usepackage{xcolor}

\usetikzlibrary{arrows.meta, positioning, calc, shapes.geometric}
\usepgfplotslibrary{fillbetween}
\usetikzlibrary{pgfplots.units}
\usepgfplotslibrary{groupplots}
\usepgfplotslibrary{colorbrewer}

\pgfplotsset{compat=1.17}
\newcommand{\panellabel}[1]{{\raggedright\large\textbf{#1}\par}\vspace{2pt}}

\definecolor{specsynth}{gray}{0.45}
\newcommand{\swatch}[1]{\textcolor{#1}{\rule{1.1ex}{1.1ex}}\,}

\newcommand{\Mvec}{\mathbf{M}}                 %
\newcommand{\mvec}{\mathbf{m}}                 %
\newcommand{\mstar}{\mathbf{m}^{\star}}        %
\newcommand{\mgt}{\mathbf{m}^{\mathrm{GT}}}    %
\newcommand{\mx}{m_x}
\newcommand{\my}{m_y}
\newcommand{\mz}{m_z}
\newcommand{\khat}{\hat{\mathbf{k}}}           %
\newcommand{\xhat}{\hat{x}}
\newcommand{\yhat}{\hat{y}}
\newcommand{\zhat}{\hat{z}}
\newcommand{\Hdem}{\mathbf{H}^{\mathrm{dem}}}  %

\newcommand{\Cp}{C^{+}}                        %
\newcommand{\Cm}{C^{-}}                        %
\newcommand{\Cpm}{C^{\pm}}
\newcommand{\sigp}{\sigma^{+}}                 %
\newcommand{\sigm}{\sigma^{-}}
\newcommand{\sigpm}{\sigma^{\pm}}
\newcommand{\Rop}{\mathcal{R}}                 %
\newcommand{\pvec}{\mathbf{p}}                 %
\newcommand{\yvec}{\mathbf{y}}                 %

\usepackage{xspace}
\newcommand{\sxz}{$xz$ series\xspace}
\newcommand{\syz}{$yz$ series\xspace}
\newcommand{\thxz}{\theta_{xz}}
\newcommand{\thyz}{\theta_{yz}}

\newcommand{\Ldata}{\mathcal{L}_{\mathrm{data}}}
\newcommand{\Ldataabs}{\mathcal{L}_{\mathrm{data}}^{\mathrm{abs}}} %
\newcommand{\Ltv}{\mathcal{L}_{\mathrm{TV}}}
\newcommand{\Rreg}{R}                          %
\newcommand{\Jobj}{\mathcal{L}}                %
\newcommand{\errm}{\varepsilon_{m}}            %

\newcommand{\Etot}{E_{\mathrm{total}}}
\newcommand{\Eex}{E^{\mathrm{ex}}}
\newcommand{\Edem}{E^{\mathrm{dem}}}
\newcommand{\Edmi}{E^{\mathrm{dmi}}}

\newcommand{\Ms}{M_{\mathrm{s}}}               %
\newcommand{\lex}{\ell_{\mathrm{ex}}}          %
\newcommand{\LD}{L_D}                          %
\newcommand{\lD}{\ell_D}                       %

\newcommand{\lamtik}{\lambda^{\mathrm{Tik}}}   %
\newcommand{\lamE}{\lambda^{E}}                %
\newcommand{\lamTV}{\lambda^{\mathrm{TV}}}     %
\newcommand{\lamtikbal}{\lamtik_{\mathrm{bal}}} %
\newcommand{\lamlow}{\lamE_{\mathrm{low}}}      %
\newcommand{\lambal}{\lamE_{\mathrm{bal}}}
\newcommand{\lamhigh}{\lamE_{\mathrm{high}}}

\newcommand{\CoZnMnComp}{\mathrm{Co}_{8}\mathrm{Zn}_{9}\mathrm{Mn}_{3}}

\begin{document}
\raggedbottom

\title{Physics-informed X-ray tomography of magnetization and material parameters}

\author{Alexander Setescak}
\email{alexander.setescak@univie.ac.at}
\affiliation{Faculty of Physics, University of Vienna, 1090 Vienna, Austria}
\affiliation{University of Vienna, Vienna Doctoral School in Physics, 1090 Vienna, Austria}

\author{Polly Mitchell}
\affiliation{Max Planck Institute for the Chemical Physics of Solids, 01187 Dresden, Germany}

\author{Luke J. R. Higgins}
\affiliation{Diamond Light Source Ltd., Harwell Science and Innovation Campus, Didcot OX11 0DE, United Kingdom}

\author{Burkhard Kaulich}
\affiliation{Diamond Light Source Ltd., Harwell Science and Innovation Campus, Didcot OX11 0DE, United Kingdom}

\author{Luke Turnbull}
\affiliation{Max Planck Institute for the Chemical Physics of Solids, 01187 Dresden, Germany}
\affiliation{Diamond Light Source Ltd., Harwell Science and Innovation Campus, Didcot OX11 0DE, United Kingdom}

\author{Rikako Yamamoto}
\affiliation{Max Planck Institute for the Chemical Physics of Solids, 01187 Dresden, Germany}
\affiliation{International Institute for Sustainability with Knotted Chiral Meta Matter (WPI-SKCM$^{2}$), Hiroshima University, Higashi-Hiroshima, Hiroshima 739-8531, Japan}

\author{Marina Raboni Ferreira}
\affiliation{Max Planck Institute for the Chemical Physics of Solids, 01187 Dresden, Germany}

\author{Daniel A. Mayoh}
\affiliation{Department of Physics, University of Warwick, Coventry CV4 7AL, United Kingdom}

\author{Geetha Balakrishnan}
\affiliation{Department of Physics, University of Warwick, Coventry CV4 7AL, United Kingdom}

\author{Anuj Chavan}
\affiliation{Faculty of Physics, University of Vienna, 1090 Vienna, Austria}

\author{Dieter Suess}
\affiliation{Faculty of Physics, University of Vienna, 1090 Vienna, Austria}

\author{Amalio Fern\'andez-Pacheco}
\affiliation{Institute of Applied Physics, TU Wien, 1040 Vienna, Austria}

\author{Manuel Guizar-Sicairos}
\affiliation{PSI Center for Photon Science, Paul Scherrer Institute, 5232 Villigen PSI, Switzerland}
\affiliation{Institute of Physics, \'Ecole Polytechnique F\'ed\'erale de Lausanne (EPFL), 1015 Lausanne, Switzerland}

\author{Claire Donnelly}
\affiliation{Max Planck Institute for the Chemical Physics of Solids, 01187 Dresden, Germany}
\affiliation{PSI Center for Photon Science, Paul Scherrer Institute, 5232 Villigen PSI, Switzerland}
\affiliation{Department of Materials, ETH Zurich, 8093 Zurich, Switzerland}

\author{Claas Abert}
\affiliation{Faculty of Physics, University of Vienna, 1090 Vienna, Austria}

\date{\today}

\begin{abstract}
Three-dimensional spin textures are stabilized by microscopic interactions within
  their host materials.
X-ray vector tomography can image these nanoscale structures, but weak magnetic contrast and a
  limited range of viewing angles often leave parts of the magnetization poorly determined.
Here we show that a reconstruction constrained by the micromagnetic energy recovers the
  three-dimensional magnetization from such data and gains sensitivity to the underlying material
  interactions.
Applied to the chiral magnet $\CoZnMnComp$, the method resolves the three-dimensional helical
  texture even from a small fraction of the dual-axis projections typically required for a
  successful tomographic reconstruction, or from a single tilt axis alone.
By testing which ratio of exchange to Dzyaloshinskii--Moriya interaction in the reconstruction
  best reproduces the measured projections, we obtain a value for the sample that differs by about
  $10\,\%$ from the literature value derived from the helical winding period.
Taken together, our findings suggest a way to reduce the time needed for vector tomography and to
  extend it to determining fundamental magnetic interactions in a material.
\end{abstract}

\maketitle

\begin{figure*}[t]
    \resizebox{\textwidth}{!}{%
        \import{figures/}{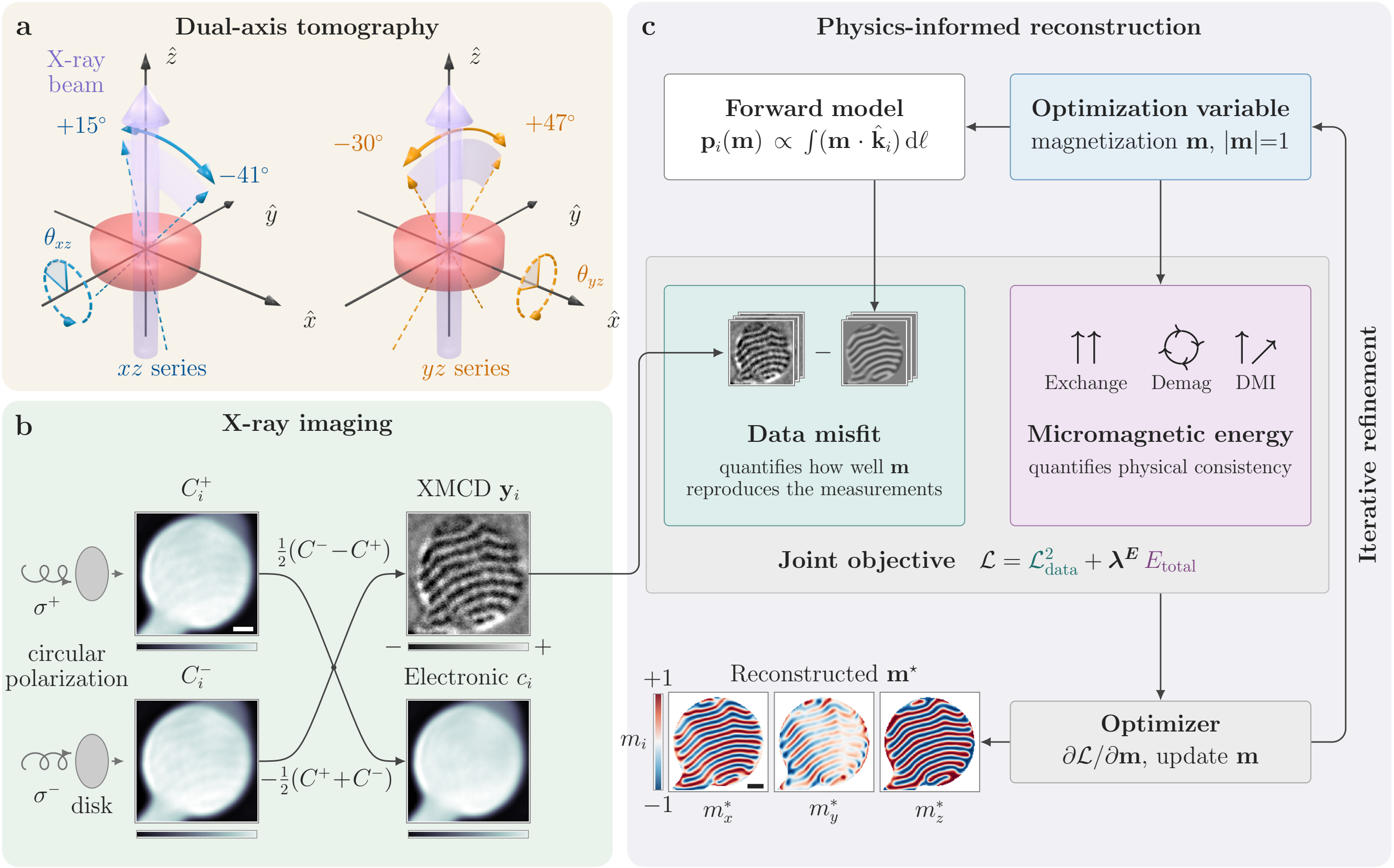}}
    \caption{\textbf{Physics-informed magnetization reconstruction from XMCD tomography.}
    \textbf{a}, Dual-axis tomography.
    Circularly polarized X-rays transmit through the sample, which we rotate about $\yhat$ by
      $\thxz \in [-41^\circ, +15^\circ]$ (\sxz) and about $\xhat$ by $\thyz \in [-30^\circ, +47^\circ]$
      (\syz).
    \textbf{b}, X-ray imaging.
    At every tilt we record transmission images at both circular polarizations $\sigpm$.
    The pair $\Cpm$ separates into the magnetic XMCD channel $y$ and the non-magnetic electronic
      channel $c$ [Eq.~\eqref{eq:cx_split}].
    \textbf{c}, The measured projections $\yvec_i$ feed into the physics-informed reconstruction, which
      minimizes the joint objective $\Jobj = \Ldata^2 + \lamE\,\Etot$ [Eqs.~\eqref{eq:loss}
      and~\eqref{eq:etot}], shown at $\lamE = 10^{14}$.
    Gradient-based optimization returns the reconstruction $\mstar$, shown by its components $m_i$ ($i = x, y, z$).
    Scale bars, \SI{200}{\nano\metre}.}
    \label{fig:setup}
\end{figure*}

Magnetic materials support a rich family of three-dimensional textures, including skyrmion
  strings~\cite{seki_direct_2022,birch_real-space_2020}, hopfions~\cite{kent_creation_2021}, and
  vortex rings~\cite{donnelly_experimental_2021}.
These textures carry a nontrivial topology that governs how they form, move, and interact, which
  makes them candidates for three-dimensional spintronic
  devices~\cite{nagaosa_topological_2013,fert_magnetic_2017}.
To understand the three-dimensional structure and behavior of these textures, we must image the
  full magnetization vector field.
X-ray magnetic vector tomography provides this capability by reconstructing the vector
  field from projections recorded with X-ray magnetic circular dichroism (XMCD)
  contrast~\cite{schutz_absorption_1987} about two rotation
  axes~\cite{donnelly_three-dimensional_2017,donnelly_tomographic_2018,hierro-rodriguez_revealing_2020}.
X-ray vector imaging has since been extended to laminography~\cite{donnelly_time-resolved_2020},
  and electron magnetic tomography has been used to reconstruct magnetic vector fields from
  holographic~\cite{wolf_unveiling_2022} or Lorentz~\cite{phatak_three-dimensional_2010} images.
However, textures that vary across the depth as well as in the plane require samples several
  hundred nanometers or even micrometers thick.
To study these systems, high-resolution imaging has been made possible with the use of hard
  X-rays~\cite{donnelly_three-dimensional_2017} and pre-edge phase contrast with coherent soft
  X-ray techniques~\cite{neethirajan_soft_2024}.
However, weak magnetic scattering can result in noisy dichroic projections.
This, combined with practical constraints on accessible measurement angles, radiation dose, and
  available synchrotron beamtime, can result in nanoscale features in the magnetization vector
  field being poorly constrained~\cite{crowther_reconstruction_1970}.

To prevent noise from distorting the reconstruction, we can spatially smooth it by penalizing
  sharp changes in the magnetization~\cite{bruckner_solving_2017,silinga_model-based_2025}.
Mathematically, this gradient penalty has the same form as the exchange interaction (stiffness $A$)
  that aligns neighboring moments in micromagnetic
  theory~\cite{brown_micromagnetics_1963,abert_micromagnetics_2019}, giving the smoothing a direct
  physical basis.
However, because the experimental resolution often approaches the scale of the magnetic features,
  this smoothing alone blurs the domain walls and helical winding we may want to image.
In micromagnetic theory, stable states are local minima of the full energy functional, of which exchange is
  only one term.
In a real magnet, these features persist because exchange competes with the demagnetizing
  interaction (scaled by the saturation magnetization $\Ms$), which favors flux closure, and in
  chiral magnets also with the Dzyaloshinskii--Moriya interaction (constant $D$), which twists
  neighboring moments.
Previous studies have applied the micromagnetic model only after the reconstruction, adjusting
  simulations by hand to reproduce the imaged
  texture~\cite{birch_real-space_2020,wolf_unveiling_2022,birch_toggle-like_2022}.
This manual tuning is time-consuming and cannot restore features that the reconstruction has
  already lost or distorted.

Here, we present an approach that avoids this loss of information by constraining the
  reconstruction itself with the micromagnetic energy, similar to our approach when recovering the
  magnetization of a thin film from stray-field magnetometry~\cite{setescak_fourier-space_2026}.
We extend X-ray magnetic vector tomography with the complete micromagnetic energy functional
  of the sample and demonstrate its two roles on a thick disk of the room-temperature chiral magnet
  $\CoZnMnComp$~\cite{tokunaga_new_2015}, imaged with dichroic soft X-ray ptychography.
First, the energy encodes our prior knowledge of the interactions in the material and helps us
  recover the magnetization vector field where a limited number of noisy projections alone is not
  enough.
Second, because the reconstruction explicitly incorporates these competing interactions, it allows
  us to test the assumed model itself.
Since material parameters are usually only determined for the bulk and do not always translate to
  a nanostructure, we identify the relevant parameters of the sample that best reproduce the
  measured projections.
This prospect of texture-driven magnetometry on individual nanostructures is relevant for a wide
  variety of magnetic material systems.

\section{\label{sec:forward}Physics-informed reconstruction from XMCD projections}
In magnetic tomography, XMCD imaging records two projections at every sample orientation, with
  opposite circular X-ray polarization $\sigpm$, giving a transmission image pair $\Cp$ and
  $\Cm$~\cite{donnelly_high-resolution_2016,donnelly_three-dimensional_2017}.
At the X-ray absorption edges of magnetic materials, the XMCD effect provides contrast that is
  sensitive to the local magnetization component along the beam propagation
  direction~\cite{schutz_absorption_1987} and reverses sign between
  $\sigp$ and $\sigm$.
We sum $\Cp$ and $\Cm$ to yield a purely electronic channel $c$ and subtract them to yield a
  magnetic (XMCD) channel $y$ (Fig.~\ref{fig:setup}b and Methods).
Because XMCD contrast is proportional to the magnetization component parallel to the beam propagation direction, each
  measured projection $\yvec_i$ records the line integral of that component along the
  beam~\cite{donnelly_three-dimensional_2017}.
In our geometry, the beam stays along $\hat{\mathbf{z}}$ while the stage rotates the sample by
  $\mathbf{R}_i$ (Fig.~\ref{fig:setup}a), so in the sample frame the beam direction is
  $\khat_i = \mathbf{R}_i^T\hat{\mathbf{z}}$~\cite{donnelly_tomographic_2018}. The forward model
  simulates the measurement of a given magnetization configuration: for normalized magnetization
  $\mvec = \Mvec/\Ms$, it computes the simulated projection $\pvec_i(\mvec)$, the counterpart of
  the measured $\yvec_i$, via the discrete Radon transform $\Rop_i$, the operator that takes these line integrals~\cite{balke_scientific_2022}:
\begin{equation}
\pvec_i(\mvec) \propto \Rop_i\!\left[\mvec\cdot\khat_i\right].
\label{eq:forward}
\end{equation}

Unlike the scalar density imaged in conventional tomography, the magnetization is a vector field
  probed only along the beam, so rotating the sample changes not only the viewing angle but also
  which component is sensed.
To reconstruct all three vector components, the acquisition therefore needs to combine two
  orthogonal tilt series (Fig.~\ref{fig:setup}a) or to use a laminography
  geometry~\cite{donnelly_time-resolved_2020}.
Tilting about $\yhat$ confines the beam to the $xz$ plane, so this series (the \sxz) is sensitive
  to $\mx$ and $\mz$ but never to $\my$.
Tilting about $\xhat$ confines the beam to the $yz$ plane, so the \syz\ is sensitive to $\my$ and
  $\mz$ but never to $\mx$.

Recovering the three-dimensional magnetization configuration requires finding a vector field whose
  forward projections reproduce the measured data (Fig.~\ref{fig:setup}c).
Even a complete dual-axis acquisition does not uniquely determine the magnetization: line integrals of $\mvec\cdot\khat$
  determine at most the divergence-free part of the
  magnetization~\cite{braun_tomographic_1991,prince_tomographic_1994}, since any internal gradient
  field that vanishes at the sample boundary integrates to zero along every beam
  path~\cite{donnelly_tomographic_2018}.
In physical terms, the magnetic volume and surface charges of the sample cannot be fully recovered
  without additional constraints.
In practice, depending on the sample geometry and the setup, projections cannot always be recorded
  over the complete angular range, leaving a so-called missing wedge of unrecorded viewing directions in
  tomography and a missing cone in laminography~\cite{holler_three-dimensional_2019}.
Moreover, weak XMCD signals make the projections particularly susceptible to noise and imaging
  artifacts across the angular range.

Despite these challenges, iterative magnetic vector
  tomography~\cite{donnelly_three-dimensional_2017,donnelly_tomographic_2018} has recovered
  three-dimensional magnetic textures by minimizing the data misfit between simulated and measured
  projections.
However, without further constraints, these limitations can leave nanoscale features poorly
  resolved and introduce artifacts into the reconstruction.
We therefore add the micromagnetic energy of the material to this data misfit and minimize the
  combined objective function $\Jobj(\mvec)$,
\begin{equation}
\Jobj(\mvec) = \Ldata^2(\mvec) + \lamE\,\Etot(\mvec),
\label{eq:loss}
\end{equation}
where the data misfit $\Ldata$, a normalized form of the error metric of magnetic vector
  tomography~\cite{donnelly_tomographic_2018}, measures how closely the simulated projections
  reproduce the $N$ measured projections $\yvec_i$,
\begin{equation}
\Ldata^2(\mvec) = \frac{1}{N}\sum_{i=1}^{N} \frac{\lVert \pvec_i(\mvec) - \yvec_i \rVert^2}{\lVert \yvec_i \rVert^2}.
\label{eq:ldata}
\end{equation}
The energy weight $\lamE$ balances this data misfit against the micromagnetic energy.

\begin{figure*}[!t]
\centering
\import{figures/}{lcurve.pgf}%
\caption{\textbf{Influence of the energy weight $\lamE$ on the reconstruction.}
\textbf{a},~L-curve of micromagnetic energy $\Etot$ versus data misfit $\Ldata$ as $\lamE$ is swept,
  marking three reconstructions with increasing weight: a low weight ($\lamlow = 10^{12}$),
  the balanced weight ($\lambal = 10^{14}$), and a high readout weight ($\lamhigh = 10^{20}$, Sec.~\ref{sec:material}).
\textbf{b},~Reconstructed magnetization components $\mx$, $\my$, and $\mz$ at these three
  operating points.
Scale bar, \SI{200}{\nano\metre}.}
\label{fig:lcurve}
\end{figure*}

The micromagnetic energy enters Eq.~\eqref{eq:loss} as the \emph{physics prior}.
This penalty increases with the energy of the magnetization state, driving the reconstruction
  toward physically plausible configurations.
It sums the exchange, demagnetizing, and bulk Dzyaloshinskii--Moriya (DMI) energies,
\begin{equation}
\Etot(\mvec) = \Eex + \Edem + \Edmi,
\label{eq:etot}
\end{equation}
  parameterized by the exchange stiffness $A$, saturation magnetization $\Ms$, and DMI constant
  $D$ [Eq.~\eqref{eq:energies}, Methods].
The micromagnetic energy is only meaningful inside the material, where the magnetization has unit
  length ($|\mvec| = 1$), whereas the surrounding vacuum carries no magnetization at all.
We therefore define a disk mask $w$ that marks which cells of the reconstruction volume belong to
  the sample, and restrict both the forward model and the energy computation to these cells.
Previous studies defined such a mask from a scalar tomogram of the non-magnetic electronic
  channel~\cite{donnelly_three-dimensional_2017}.
For our flat disk, however, the missing wedge elongates this tomogram along $z$ (Methods).
We instead take the disk outline from the electronic channel of the face-on projection and
  extend it through the fabricated \SI{300}{\nano\metre} thickness.
We minimize $\Jobj$ using gradient-based optimization, with automatic differentiation supplying
  the gradients~\cite{abert_neuralmag_2025}.

\section{\label{sec:results}Reconstructing the magnetic state}
To demonstrate our physics-informed reconstruction, we apply it to tomographic data measured on a
  \SI{1.1}{\micro\metre}-diameter, \SI{300}{\nano\metre}-thick $\CoZnMnComp$ disk to recover its
  three-dimensional magnetization configuration.
$\CoZnMnComp$ belongs to the Co--Zn--Mn family of room-temperature
  helimagnets~\cite{tokunaga_new_2015}, which host skyrmions~\cite{tokunaga_new_2015}, Bloch
  points~\cite{birch_toggle-like_2022}, and hopfions~\cite{zhu_magnetic_2026}.
We image the disk in the remanent state, where it hosts a helical configuration.
The XMCD phase projections were measured \SI{3}{\electronvolt} below the Co $L_3$ absorption edge
  (\SI{778.1}{\electronvolt}) using dichroic soft X-ray ptychography, a coherent diffractive imaging method that
  recovers the phase shift of the transmitted X-rays.
At this pre-edge energy, the X-rays are only weakly attenuated while the magnetic phase contrast
  persists, which allows imaging of samples up to an order of magnitude thicker than at the absorption edge~\cite{neethirajan_soft_2024}.
The two tilt series cover $\thxz \in [-41^\circ, +15^\circ]$ ($29$ projections, \sxz) and
  $\thyz \in [-30^\circ, +47^\circ]$ ($38$ projections, \syz) with a pixel size of about
  \SI{10}{\nano\metre}.
To reconstruct the magnetization with Eq.~\eqref{eq:loss}, we choose the energy weight $\lamE$ and
  specify the material parameters.
We use the literature values of $A$ and $\Ms$~\cite{birch_toggle-like_2022} and
  $D = \SI{0.66e-3}{\joule\per\metre\squared}$, the DMI constant that best reproduces the measured
  projections (Sec.~\ref{sec:material}, Extended Data Table~\ref{tab:params}).

We first examine how the energy weight $\lamE$ influences the reconstructed state by sweeping
  $\lamE$ across several orders of magnitude and plotting the micromagnetic energy $\Etot$ against the data
  misfit $\Ldata$ in an L-curve (Fig.~\ref{fig:lcurve}a), similar to how regularization weights are
  determined in inverse problems~\cite{hansen_use_1993,calvetti_tikhonov_2000}.
At $\lamlow$, the physics prior is too weak to constrain the reconstruction, which is nearly
  identical to an unconstrained one ($\lamE = 0$, Extended Data Fig.~\ref{edf:synth}a) and shows a
  limitation common to magnetic vector tomography: because the X-ray beam stays close to $\zhat$
  over the restricted tilt range, the measurement strongly constrains the out-of-plane stripes in
  $\mz$, whereas the weakly probed in-plane components $\mx$ and $\my$ remain much noisier and
  less resolved (Fig.~\ref{fig:lcurve}b, top row).
At $\lamhigh$, the energy term pulls the reconstruction toward a low-energy state at the expense of
  data agreement: the $\mz$ stripes differ from both the $\lamlow$ reconstruction and the measured
  projections (Fig.~\ref{fig:lcurve}b, bottom row).
To select the balanced weight $\lambal$, we increase $\lamE$ up to the bend of the L-curve
  ($\lambal = 10^{14}$), just before the reconstruction begins to clearly deviate from the measurements
  ($\Ldata = 0.83$ versus $0.78$).
At this weight, $\mz$ keeps the stripe pattern of the $\lamlow$ reconstruction and of the measured
  projections (Extended Data Fig.~\ref{edf:residuals}), and $\mx$ and $\my$ form a coherent
  three-dimensional helical texture (Fig.~\ref{fig:lcurve}b, middle row).

However, the true magnetization of the measured disk is unknown, and the data misfit alone cannot
  show whether the reconstructed $\mx$ and $\my$ are correct.
We therefore test the method on a synthetic disk whose magnetization $\mgt$ we
  know~\cite{donnelly_tomographic_2018}.
To obtain a realistic $\mgt$, we start from the unconstrained reconstruction ($\lamE = 0$) of the
  measured disk and relax it into an energy minimum with the Landau--Lifshitz--Gilbert
  equation~\cite{landau_theory_1935,gilbert_phenomenological_2004} in
  NeuralMag~\cite{abert_neuralmag_2025} (Extended Data Fig.~\ref{edf:synth}a).
We then simulate the experiment on this disk: we compute its projections at the measured tilt
  angles, add noise matched to the measurement, and reconstruct the magnetization from these
  synthetic projections with the same physics-informed protocol.
Comparing the reconstruction $\mstar$ with the known $\mgt$ gives the relative error
  $\errm = \lVert \mstar - \mgt \rVert / \lVert \mgt \rVert$.
Without the physics prior ($\lamE = 0$), the reconstruction recovers the $\mz$ stripes but fails to
  resolve the in-plane components, resulting in a large overall error of $\errm = 0.74$ (Extended
  Data Fig.~\ref{edf:synth}a,e).
Including the physics prior at the balanced weight $\lambal$ accurately recovers all three
  components and reduces the error to $\errm = 0.11$.

Finally, because the material parameters supplied to the physics-based reconstruction are often
  only approximately
  known in experiment, we use the synthetic disk to test how sensitive the reconstructed state is
  to parameter mismatch.
When we reconstruct the full dataset at $\lambal$ while offsetting the input $D$ or $\Ms$ by
  $\pm30\,\%$, the error $\errm$ increases from $0.11$ to at most $0.15$, whereas offsetting $A$
  by $+30\,\%$ and $-30\,\%$ yields $\errm = 0.11$ and $0.23$, respectively (Extended Data
  Fig.~\ref{edf:synth}f).
The error rises most when $A$ is underestimated, because a weaker exchange stiffness smooths the
  measurement noise less.
Apart from this case, parameter offsets of $\pm30\,\%$ leave the reconstructed state largely
  unchanged, so normal uncertainty in the material parameters of the sample is tolerable.

\section{\label{sec:sparse}Reconstruction from sparse and single-axis data}
\begin{figure}[!t]
\centering
\import{figures/}{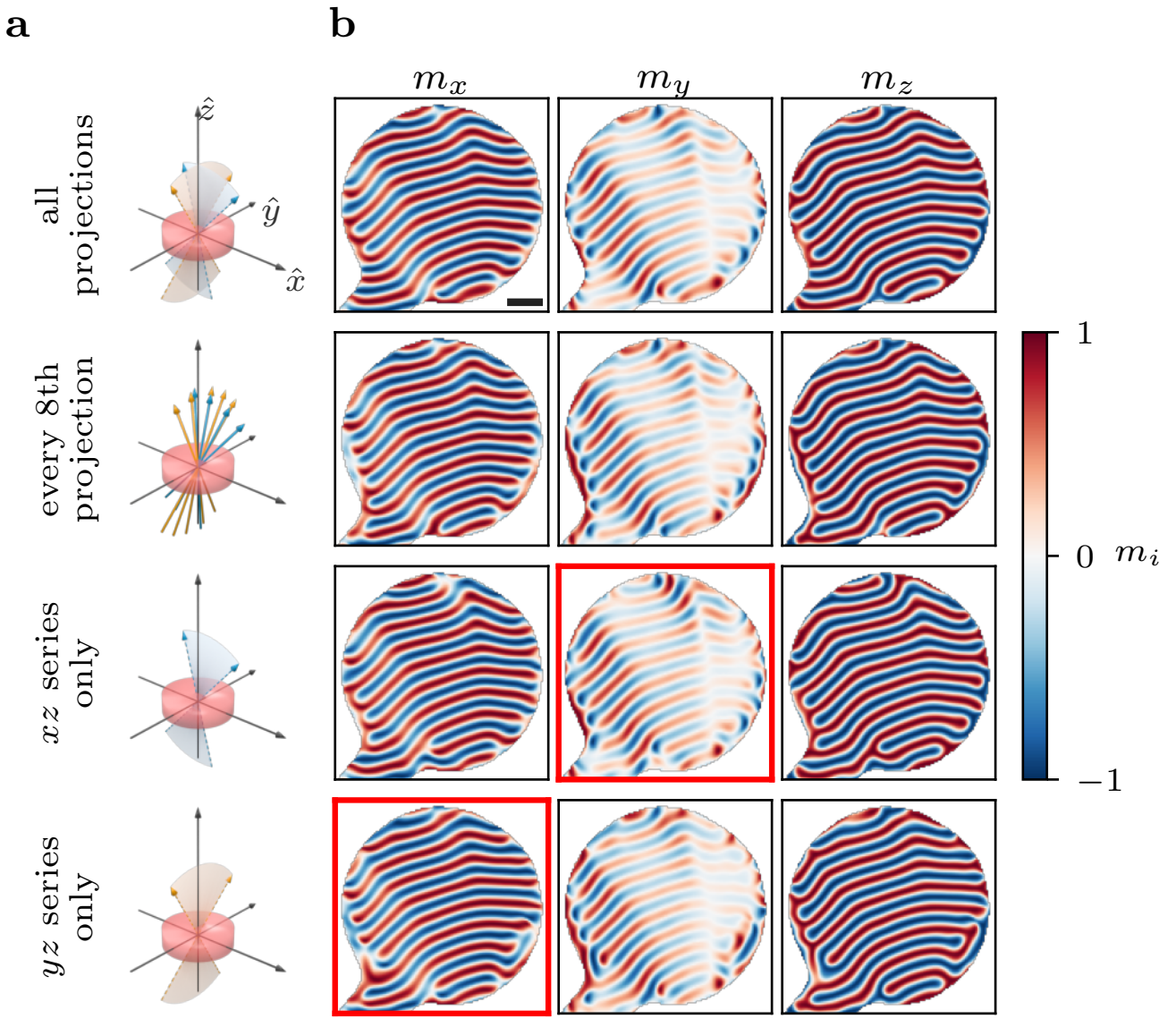}
\caption{\textbf{Reconstructions from sparse and single-axis projection data.}
One row per acquisition condition: the full dual-axis measurement combining both tilt series
  ($67$ projections), a sparse dual-axis acquisition retaining only every eighth projection of each
  series ($9$), and the two single-axis acquisitions, \sxz\ alone ($29$) and \syz\ alone ($38$).
\textbf{a},~Acquisition in the geometry of Fig.~\ref{fig:setup}a: the beam directions swept by
  the \sxz\ (blue) and by the \syz\ (orange), or the nine retained beam directions.
\textbf{b},~Disk-plane $\mx$, $\my$, $\mz$ cross sections, all reconstructed with
  the physics prior at $\lamE = 10^{15}$.
Red frames mark the component that the single series does not measure: $\my$ for the \sxz\ and
  $\mx$ for the \syz.
Scale bar, \SI{200}{\nano\metre}.}
\label{fig:projthin}
\end{figure}

Time-resolved and field-dependent studies require a complete dual-axis tilt series for every time
  step or field value.
Tomograms at successive time steps have been recorded~\cite{donnelly_time-resolved_2020}, but
  faster acquisitions would allow for better temporal resolution and a wider range of phenomena and
  experimental parameters to be explored.
To maximize the efficiency of these measurements, we test whether the physics prior can recover
  the magnetization from reduced projection data, either by sparse angular sampling or by omitting
  an entire tilt series.
We reconstruct the disk under four acquisition conditions (Fig.~\ref{fig:projthin}): the full
  dual-axis measurement, a sparse dataset with only every eighth projection of each tilt series
  ($9$ projections), and either tilt series alone (\sxz\ or \syz).
All four use $\lamE = 10^{15}$, ten times the balanced weight $\lambal$, to compensate
  for the reduced number of projections.
Both the sparse and the single-axis reconstructions reproduce the chiral stripe pattern of the
  full dataset (Fig.~\ref{fig:projthin}b).
A single tilt series rotates the sample about one axis, so the magnetization component along that
  axis never aligns with the beam and goes unmeasured (Sec.~\ref{sec:forward}).
The physics prior fills it in: the micromagnetic energy couples the unmeasured component to the two
  measured ones, so all three come out consistent (red-framed panels in Fig.~\ref{fig:projthin}b).
The two single-axis reconstructions differ mainly at the edge of the sample, where the disk mask
  fits worst, and noise differences between the two series slightly change the stripe pattern
  (Extended Data Fig.~\ref{edf:residuals}).

However, agreement between the reduced and full reconstructions of the measured disk does not show
  that the unmeasured component is correct.
We therefore repeat the four acquisitions on the synthetic disk, whose magnetization is known.
At the weight of Fig.~\ref{fig:projthin} ($\lamE = 10^{15}$), the reconstruction recovers the
  three-dimensional helical winding in all three components for every acquisition, with $\errm$
  between $0.08$ and $0.16$ (Extended Data Fig.~\ref{edf:synth}c), compared with $0.74$ for the
  complete dataset without the physics prior.
Consequently, a small fraction of the projections, or a single tilt series alone, suffices to
  recover the three-dimensional texture, so that vector tomography experiments become faster and
  can cover more experimental conditions.

\section{\label{sec:material}Reading material parameters from the data misfit}
\begin{figure*}[t]
\centering
\import{figures/}{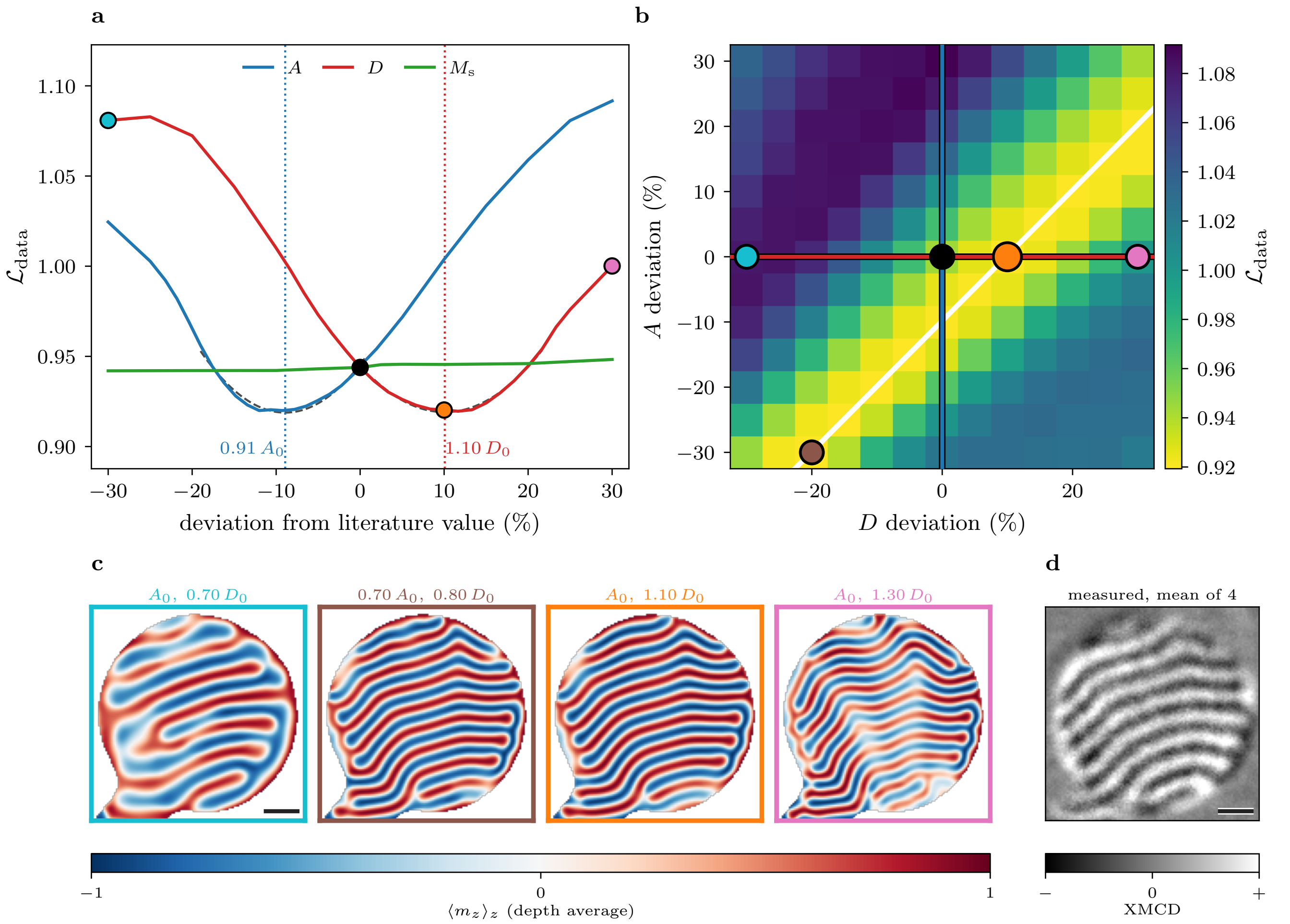}%
\caption{\textbf{Sensitivity of the data misfit to the material parameters.}
\textbf{a},~Data misfit $\Ldata$ at $\lamE = 10^{20}$ when one parameter is varied, with the other
  two fixed at the literature values $A_0$, $D_0$, $\Ms$~\cite{birch_toggle-like_2022}.
The misfit stays flat in $\Ms$, whereas $A$ and $D$ trace two wells with minima on either side of
  the literature point, $9\,\%$ below $A_0$ and $10\,\%$ above $D_0$ (dotted lines, the vertices
  of the dashed parabolas).
\textbf{b},~Data misfit over a $13\times13$ grid in $A$ and $D$ (deviation from $A_0$, $D_0$) at
  $\lamE = 10^{20}$.
The misfit is low along a valley of constant $A/D$.
The white line is the fitted trough of that valley, the red and blue lines are the $A = A_0$ and
  $D = D_0$ cuts of \textbf{a}.
\textbf{c},~Depth-averaged $\mz$ of the four reconstructions, with $A$ and $D$ labeled relative to
  $A_0$, $D_0$.
\emph{Along} the valley: the minimum of the $A = A_0$ cut (orange) and the on-valley
  cell farthest from it (brown).
\emph{Across} the valley: the low-energy states at $A = A_0$ on the grid edges, with
  $D = 0.70\,D_0$ and $1.30\,D_0$ (cyan and pink).
Markers of the same colors mark these reconstructions in \textbf{b}, and those at $A = A_0$ in
  \textbf{a}. Black marks the literature values.
\textbf{d},~Mean of the four measured XMCD projections within $2^\circ$ of normal incidence
  ($\thxz = \pm1^\circ$ and $\thyz = \pm1^\circ$), for comparison with the textures in \textbf{c}.
Scale bars, \SI{200}{\nano\metre}.}
\label{fig:adgrid}
\end{figure*}

The magnetic texture of the disk forms through the competition between the micromagnetic
  interactions, so the measured texture should contain information about the material
  parameters.
Here, we explore whether the tomographic reconstruction itself can determine the material
  parameters of the sample.
However, varying the assumed parameters has little effect at the balanced weight $\lambal$, where
  the reconstruction is primarily guided by the measurement data (Sec.~\ref{sec:results}).
To increase the sensitivity of the reconstruction to the material parameters, we instead raise
  $\lamE$ to a high weight
  ($\lamhigh$ in Fig.~\ref{fig:lcurve}b).
In this regime, the physics prior dominates the objective function, forcing $\mvec$ toward a low-energy state 
  dictated by the assumed parameters.
If those parameters differ from the sample's true constants, the resulting texture does not match
  the measured projections and the data misfit $\Ldata$ rises.
Consequently, we can find the best-matching parameters by reconstructing across candidate values at
  $\lamhigh$ and looking for the minimum in $\Ldata$.

We first vary one parameter at a time around the literature values
  $A_0 = \SI{5.73}{\pico\joule\per\metre}$, $D_0 = \SI{0.60e-3}{\joule\per\metre\squared}$, and
  $\Ms = \SI{461}{\kilo\ampere\per\metre}$~\cite{birch_toggle-like_2022}, performing a separate
  tomographic reconstruction of the disk at $\lamE = 10^{20}$ for each value (Fig.~\ref{fig:adgrid}a).
Across a $\pm30\,\%$ range in $\Ms$, $\Ldata$ changes by less than $1\,\%$, showing that the data
  misfit is relatively insensitive to the saturation magnetization.
In contrast, $A$ and $D$ form two wells of equal depth with minima at $0.91\,A_0$ and
  $1.10\,D_0$.
Because a $9\,\%$ decrease in $A$ and a $10\,\%$ increase in $D$ change the ratio $A/D$ by the same
  factor, these mirrored minima indicate that we are only sensitive to the ratio of $A/D$.
To test this across both parameters simultaneously, we reconstruct the disk over a two-dimensional
  $(A, D)$ grid (Fig.~\ref{fig:adgrid}b).
The misfit forms a continuous valley along a line of constant $A/D$, confirming that the
  tomographic data fix only the relative strength of exchange and DMI.

The constant-ratio valley reflects the competing roles of exchange and DMI in the micromagnetic
  energy, where $A/D$ sets the equilibrium winding period of the helix (given by the pitch
  $\LD = 4\pi A/D$ in an unbounded crystal)~\cite{karube_metastable_2020}.
Along the valley floor, parameter pairs with widely different absolute values therefore converge to
  nearly identical stripe textures with the same period (Fig.~\ref{fig:adgrid}c, orange and brown).
In contrast, when we step off the valley by changing $D$ by $\mp30\,\%$ at fixed $A = A_0$, the
  stripe period shifts noticeably: $D = \SI{0.42e-3}{\joule\per\metre\squared}$ (cyan) widens the
  stripes, whereas $D = \SI{0.78e-3}{\joule\per\metre\squared}$ (pink) tightens the winding.
Only parameter pairs along the valley floor reproduce the stripe period seen in the measured
  face-on projections (Fig.~\ref{fig:adgrid}d).

With $A$ fixed at $A_0$, the minimum at $1.10\,D_0$ gives $D = \SI{0.66e-3}{\joule\per\metre\squared}$
  (Fig.~\ref{fig:adgrid}a).
This well is shallow: $\Ldata$ rises by less than $1\,\%$ within
  $\SI{0.03e-3}{\joule\per\metre\squared}$ of the optimum.
Even so, its minimum stays at the same $D$ for every energy weight from $\lamE = 10^{16}$ to
  $10^{20}$ (Extended Data Fig.~\ref{edf:dsens}a).
Only at lower weights does the well flatten and shift toward smaller $D$, before disappearing
  near the balanced weight $\lambal = 10^{14}$.

Our readout lies $10\,\%$ above the literature value $D_0$, although our disk winds with the same
  $\SI{120}{\nano\metre}$ period (Fig.~\ref{fig:adgrid}d) as the nanowire from which $D_0$ was
  derived~\cite{birch_toggle-like_2022}.
The literature value follows from inserting this period into the bulk relation $D = 4\pi A_0/L$,
  which describes a helix in an unbounded crystal, whereas our readout models the full
  micromagnetic energy of the confined disk.
We test this difference with the synthetic disk, whose $D$ is known: relaxed at $A = A_0$ and
  $D = \SI{0.66e-3}{\joule\per\metre\squared}$ (Methods), it forms stripes with a
  $\SI{122}{\nano\metre}$ period, longer than the $\SI{109}{\nano\metre}$ that the bulk relation
  predicts.
Applied to this period, the bulk relation underestimates $D$ by about $10\,\%$, as in the measured
  disk.
Our readout, in contrast, returns the $D$ used to relax the synthetic disk from its simulated
  projections (Extended Data Fig.~\ref{edf:synth}f).

Even at the optimal $D$, the high-weight reconstruction ($\lamhigh$) has a larger data misfit
  than the balanced reconstruction ($\Ldata = 0.93$ versus $0.83$, Fig.~\ref{fig:lcurve}a).
In a real sample, however, pinning sites, local defects, or focused-ion-beam damage can distort
  the texture in ways that our energy model does not capture, increasing the data misfit at
  $\lamhigh$.
In our disk these local distortions are small enough that the overall texture still lies close to
  a minimum of Eq.~\eqref{eq:etot}, allowing a clear dip in $\Ldata$.
If strong pinning instead traps a sample in a metastable state that Eq.~\eqref{eq:etot} cannot
  reproduce, the high-weight reconstruction converges to a state far from the measured texture for
  every parameter set, leaving $\Ldata$ uniformly high and preventing a readout.

\section{\label{sec:conclusions}Conclusions}
We have presented an approach to imaging nanoscale magnetic textures that incorporates the
  micromagnetic energy into the objective function, with a single weight $\lamE$ selecting between
  two operating regimes of the same reconstruction.
In the balanced regime the prior helps recover the texture from noisy, incomplete measurements, and
  the result tolerates the typical uncertainty of the assumed material parameters.
A larger weight on the micromagnetic energy switches to the second regime, in which the data
  misfit can be used to read out the parameters that shape the texture.
Moreover, a parameter's impact on the reconstruction matches its sensitivity in the readout:
  parameters that barely affect the texture cannot be extracted, but they also barely change the
  reconstructed state, whereas parameters that strongly shape the texture can be determined
  directly from the misfit.
In the helimagnetic $\CoZnMnComp$ disk, the reconstruction is relatively insensitive to $\Ms$,
  whereas the ratio $A/D$ sets the period of the helical winding.
At the literature exchange stiffness $A_0$, the readout returns
  $D = \SI{0.66e-3}{\joule\per\metre\squared}$, $10\,\%$ above $D_0$, although the disk winds with
  the $\SI{120}{\nano\metre}$ period from which $D_0$ was derived.
The bulk relation behind $D_0$ describes an unbounded crystal and underestimates $D$ in the
  confined disk.
The ratio $A/D$ therefore has to be measured in the nanostructure itself.

The sparse-data recovery and the parameter readout together suggest a strategy for experimental
  design.
Dense dual-axis tilt series, typically required for high-resolution three-dimensional imaging, take
  a significant amount of beamtime in vector tomography.
With the physics prior, far fewer projections are required, and even a single tilt axis can provide
  a robust reconstruction.
Every projection a reconstruction can spare buys beamtime for additional specimens, magnetic
  fields, or time points within one allocation, and thus for more complex experiments such as in
  situ studies.
A single well-sampled measurement of a prepared reference state, such as a helical texture whose
  period pins the constants, could then validate the energy model and fix the material parameters,
  while subsequent conditions are imaged with sparse acquisitions.

We demonstrate this physics-informed approach with XMCD tomography, but automatic differentiation in
  micromagnetic software~\cite{abert_neuralmag_2025,bruckner_magnumnp_2023} allows the physics
  prior to be paired with any differentiable forward model and extended with further interactions
  of the host material.
The same approach can therefore be extended to other magnetic imaging modalities, including
  electron holographic and Lorentz
  tomography~\cite{wolf_unveiling_2022,phatak_three-dimensional_2010} and
  laminography~\cite{donnelly_time-resolved_2020}.

\section*{Methods}

\subsection*{Data acquisition and preprocessing}

\paragraph{Sample.}
The $\CoZnMnComp$ disk of Extended Data Table~\ref{tab:params} sits on a focused-ion-beam
  support wedge.
The disk was cut from a single crystal grown at the University of Warwick by the Bridgman
  method~\cite{neethirajan_soft_2024}.
The table gives its acquisition and discretization.
The material parameters $\Ms$, $A_0$, and $D_0$ are those of Ref.~\cite{birch_toggle-like_2022}
  for the same compound: $\Ms$ from dc magnetization data, and $A_0$ with $D_0$ chosen there to
  reproduce the helical period $\LD = 4\pi A_0/D_0 = \SI{120}{\nano\metre}$ imaged in a
  \SI{200}{\nano\metre} thick nanowire.

\paragraph{X-ray data acquisition.}
We acquired the projections at room temperature using dichroic soft X-ray
  ptychography~\cite{donnelly_high-resolution_2016,donnelly_three-dimensional_2017} on the I08-1
  ptychography end station of Diamond Light Source under proposal MG37535\nobreakdash-2.
An APPLE-II undulator produced circularly polarized X-rays of either helicity, set by its gap and
  phase, and the Au \SI{600}{lines\per\milli\metre} grating of the SX-700 plane-grating
  monochromator~\cite{yang_design_2017} tuned them \SI{3}{\electronvolt} below the Co $L_3$ edge
  (\SI{778.1}{\electronvolt}).
A \SI{10}{\micro\metre} pinhole defined the secondary source, and a Fresnel zone plate with a
  diameter of \SI{333}{\micro\metre}, an outermost zone width of \SI{70}{\nano\metre}, and a mean
  focal length of \SI{14.156}{\milli\metre} between 770 and \SI{807}{\electronvolt} focused the
  beam.
The sample sat on an L-bracket on an SR-5714C rotation stage (SmarAct GmbH), about
  \SI{50}{\micro\metre} downstream of the focus, and we scanned it through the beam along a spiral
  path to avoid grid artifacts.
An in-vacuum Axis-SXRF-2020EUV sCMOS detector (Axis Photonique Inc.)\ recorded the diffraction
  patterns.
We measured the two polarizations separately as $\Cp$ and $\Cm$, each yielding a series of
  diffraction patterns.
At the absorption peak, resonant attenuation would block transmission through the disk, whereas at
  this pre-edge energy the transmission remains high while the magnetic phase contrast
  persists~\cite{neethirajan_soft_2024}.
We reconstructed the complex-valued object projections (amplitude and phase contrast images) and
  the probe with PtyPy~\cite{enders_computational_2016} after cropping the
  diffraction patterns to $1024 \times 1024$ pixels and binning them to $512 \times 512$ pixels,
  running 1000 iterations of the relaxed averaged alternating reflections (RAAR) engine,
  200 iterations of maximum-likelihood refinement, and another 500 RAAR iterations.

\paragraph{Preprocessing.}
Each complex projection carries an absorptive channel (the log amplitude) and a dispersive channel
  (the phase).
In either channel, we combine the projections of the two polarizations to isolate the electronic
  term $c$ and the XMCD term $y$,
\begin{equation}
c = -\tfrac{1}{2}\!\left(\Cp+\Cm\right),
\qquad
y = \tfrac{1}{2}\!\left(\Cm-\Cp\right) .
\label{eq:cx_split}
\end{equation}
Both encode the same projected moment, but at the pre-edge energy the phase XMCD has the higher
  signal-to-noise ratio, and therefore is selected as the reconstruction input.
Starting from a coarse pre-alignment of the projections, we fit an ellipse to
  the projected disk outline in every frame and shift its center onto the rotation
  axis~\cite{odstrcil_alignment_2019}, register the exposures of each angle (both polarizations and
  any repeats) against each other on the electronic channel before averaging them, and apply the
  composed shift in a single cubic interpolation per exposure.
We then remove the phase ramp of every projection by fitting a plane to the phase over the
  air region, detected by an Otsu threshold~\cite{otsu_threshold_1979} on the transmission
  amplitude, and subtracting it.
For the electronic channel, we subtract from $c$ its median outside the disk support in every
  frame, which normalizes the transmitted amplitude to the background.

\paragraph{Tilt geometry.}
The stage angles enter the forward model through one rotation per series about its axis.
We calibrate the zero point $\theta_0$ of each series from the same disk outlines.
The disk outline is a circle when the disk faces the beam and an ellipse that narrows as the
  disk tilts away.
We fit the aspect ratio of this ellipse in all frames of a series at once with
  $a(\theta) = s\,[\cos(\theta + \theta_0) + k\,|\sin(\theta + \theta_0)|]$, where $\theta$ is the
  stage angle recorded with each frame in the raw data.
The fit has three free parameters: the zero point $\theta_0$, a scale $s$, and $k$, which accounts
  for the blurred, partly absorbing rim.
We apply $\theta_0$ rounded to the nearest degree, the precision of the stage labels
  (Extended Data Table~\ref{tab:params}).

\subsection*{Magnetic reconstruction framework}

\paragraph{Loss functional.}
Both the physics prior and the Tikhonov reconstruction minimize a squared data misfit plus one
  regularizer $\Rreg$ with weight $\lambda$,
\begin{equation}
\Jobj(\mvec) = \Ldata^2(\mvec) + \lambda\,\Rreg(\mvec).
\label{eq:loss_generic}
\end{equation}
In the data term of Eq.~\eqref{eq:ldata}, $\pvec_i$ is the forward model of
  Eq.~\eqref{eq:forward_methods}, $\yvec_i$ the measured XMCD projection, and the norms run over
  the projection pixels of projection $i$.
Each residual is normalized by that projection's own measurement power, so every view carries equal
  weight.
The residual is always evaluated on the full projection frame: the disk mask acts on the magnetization
  field, never on the data residual.

\paragraph{Forward model.}
The Tikhonov reconstructions use an isotropic $120\times120\times120$ tomography grid with a cell
  spacing of $\approx\SI{9.96}{\nano\metre}$ (one cell per projection pixel across the central projection
  window).
The physics reconstructions instead run on a box that just encloses the disk (Extended Data
  Table~\ref{tab:params}).
In the lab frame the beam runs along $\hat{\mathbf{z}}$, the projection pixel $(x, y)$ spans the
  transverse plane, and the stage rotation $\mathbf{R}_i$ carries a sample point $\mathbf{r}$ to the
  lab position $\mathbf{R}_i\mathbf{r}$ (Fig.~\ref{fig:setup}a).
Equation~\eqref{eq:forward} written out with the magnetization scale $\gamma$ and the disk
  mask $w$ gives the value $p_i(x, y)$ of the projection $\pvec_i(\mvec)$ at projection
  pixel $(x, y)$ as the line integral along the beam,
\begin{equation}
p_i(x, y) = \gamma \int \bigl[w\,\mvec\cdot\khat_i\bigr]\!\bigl(\mathbf{R}_i^{T}\mathbf{r}\bigr)\,\mathrm{d}z,
  \qquad \mathbf{r} = (x, y, z),
\label{eq:forward_methods}
\end{equation}
where the sample-frame field is evaluated at the sample point the rotation carries to $\mathbf{r}$,
  and $\gamma$ is the amplitude the projector sees at every cell inside the mask: under the
  unit-norm constraint $|\mvec| = 1$ the projected magnetization has $|\gamma\mvec| = \gamma$
  everywhere in the material.
It collects $\Ms$, the magnetic scattering factor, the atomic density and the cell size into one
  constant~\cite{donnelly_three-dimensional_2017}, which we read from the Tikhonov reconstructions
  ('Magnetization-scale calibration' below).
We evaluate this integral on the cell grid with the differentiable projector of
  SCICO~\cite{balke_scientific_2022}.

\paragraph{Tikhonov regularization.}
The Tikhonov reconstruction treats $\mvec$ as an unconstrained Cartesian vector field inside the
  disk mask $w$, without a magnetization norm constraint: it replaces $\mvec$ by $w\,\mvec$ in
  the forward model, so cells outside the mask carry no magnetization and the penalty
  $\Rreg(\mvec) = \lVert\mvec\rVert^2$ of Eq.~\eqref{eq:loss_generic} sums $|\mvec|^2$ over the disk.
We minimize the resulting objective with the L-BFGS implementation in
  JAXopt~\cite{blondel_efficient_2022}, relying on its zoom line search to set the step size.
On the measured data we use $\lamtikbal = 1$ (Extended Data Fig.~\ref{edf:solnorm}a).
An unmasked variant reconstructs the field freely on the full box.

\paragraph{Micromagnetic energy regularization.}
The physics reconstruction sets $\Rreg = \Etot$ and $\lambda = \lamE$, the plain micromagnetic
  energy of Eq.~\eqref{eq:etot} in joules, so $\lamE$ carries the inverse units of energy.
The unit-norm constraint $|\mvec|=1$ needs the mask $w$ to separate material from vacuum
  and the scale $\gamma$ of Eq.~\eqref{eq:forward_methods} to match the measured amplitude
  (Extended Data Table~\ref{tab:params} gives the bounding box).
With exchange stiffness $A$, saturation magnetization $\Ms$, demagnetizing field $\Hdem$, and bulk
  Dzyaloshinskii--Moriya constant $D$, its terms take their standard
  form~\cite{abert_micromagnetics_2019},
\begin{align}
\Eex  &= \int_\Omega A\,\lvert\nabla\mvec\rvert^2\,\mathrm{d}V, \nonumber\\
\Edem &= -\tfrac{1}{2}\mu_0 \Ms \!\int_\Omega \mvec\cdot\Hdem\,\mathrm{d}V, \nonumber\\
\Edmi &= \int_\Omega D\,\mvec\cdot(\nabla\times\mvec)\,\mathrm{d}V.
\label{eq:energies}
\end{align}
We compute the energy with NeuralMag~\cite{abert_neuralmag_2025} on its JAX backend.
It evaluates the local terms by finite differences and the demagnetizing field by FFT convolution.

The energy must be evaluated on a grid that resolves the characteristic length scales of the
  material, the exchange length $\lex = \sqrt{2A/(\mu_0 \Ms^2)} \approx \SI{6.6}{\nano\metre}$ and
  the DMI length $\lD = 2A/D \approx \SI{17.4}{\nano\metre}$.
The tomography cell of $\approx\SI{9.96}{\nano\metre}$, the projection pixel, bounds the spatial
  resolution and samples the $\SI{120}{\nano\metre}$ helical period with $\approx 12$ cells, but it
  sits above $\lex$, so
  we refine it by a factor of two and evaluate the energy on the resulting $\SI{4.98}{\nano\metre}$
  physics grid ($\approx 24$ cells per period), below both micromagnetic length scales.
We reconstruct the magnetization on this finer grid and block-average it onto the tomography grid
  before projection.
\paragraph{Optimization details.}
Riemannian Adam~\cite{becigneul_riemannian_2019}, the manifold generalization of
  Adam~\cite{kingma_adam_2017}, keeps $|\mvec|=1$ exactly at every step (custom JAX
  implementation).
We run at a constant learning rate of $0.4$ with the standard coefficients ($\beta_1 = 0.9$,
  $\beta_2 = 0.999$, $\epsilon = 10^{-8}$) for 300 to 400 iterations from a cold start at $\mvec = 0$.
The zero vector lies outside the unit sphere, so the first step is a Euclidean gradient step whose
  normalization places every cell on the sphere, and all later steps are Riemannian.

Each physics reconstruction fits on one 24\,GB MIG slice of an NVIDIA RTX PRO 6000 Blackwell GPU
  in single precision.
The energy, the projector, and the optimizer update fuse into one kernel under just-in-time
  compilation~\cite{bradbury_jax_2018}, which costs about $55\,\mathrm{s}$ once per configuration.
On the full GPU, a compiled 400-iteration reconstruction then takes $13\,\mathrm{s}$, and
  $\lambda$ enters the kernel as a runtime argument, so the nine-point L-curve sweep
  ($\lamE = 10^{12}$ to $10^{20}$ in decade steps) reuses one compilation and completes in about
  three minutes.

\paragraph{Disk mask.}
We obtain the in-plane disk outline by Otsu thresholding~\cite{otsu_threshold_1979} the electronic
  channel of the zero-tilt projection (Fig.~\ref{fig:setup}b).
The outline includes the short neck that joins the disk to the support wedge, so the reconstructed
  texture continues into it.
We set the mask to this uniform outline across 30 cells in the $z$-direction, matching the
  fabricated $\SI{300}{\nano\metre}$ sample thickness (Extended Data Table~\ref{tab:params}).
A separate scalar tomography of the electronic channel confirms this mask (Extended Data Fig.~\ref{edf:charge}).

\paragraph{Electronic-channel tomography.}
We validate the disk mask with a scalar tomography of the electronic channel $c$ of
  Eq.~\eqref{eq:cx_split}, taken from the log amplitude.
We reconstruct a non-negative density $\rho$ on a free $120^3$ box of the tomography grid, without
  the mask, by minimizing
\begin{equation}
\Jobj(\rho) = \left(\Ldataabs\right)^2(\rho) + \lamTV\,\Ltv(\rho),
\label{eq:loss_tv}
\end{equation}
where the squared misfit averages the relative residual over all $N$ projections $\mathbf{c}_i$,
\begin{equation}
\left(\Ldataabs\right)^2(\rho) = \frac{1}{N}\sum_{i=1}^{N} \frac{\lVert \Rop_i\rho - \mathbf{c}_i \rVert^2}{\lVert \mathbf{c}_i \rVert^2},
\label{eq:ldata_abs}
\end{equation}
the electronic-channel counterpart of Eq.~\eqref{eq:ldata}.
Here $\Rop_i$ is the scalar line integral of Eq.~\eqref{eq:forward_methods} along the beam
  of projection $i$, and $\Ltv$ is the anisotropic total variation, the absolute differences
  between neighboring cells averaged over the cells of the box.
Total variation favors piecewise-constant densities with sharp edges, the form of a uniform disk in air.
We enforce $\rho \geq 0$ through $\rho = u^2$ and run 300 L-BFGS iterations from a uniform start
  for each $\lamTV$ from $10^{-2}$ to $10^{2}$ in decade steps.
The root $\Ldataabs$ is the misfit axis of the L-curve (Extended Data Fig.~\ref{edf:charge}).

\paragraph{Magnetization-scale calibration.}
The magnetization scale $\gamma$ of Eq.~\eqref{eq:forward_methods} matches the unit-norm
  magnetization to the measured XMCD amplitude~\cite{donnelly_three-dimensional_2017}.
We read it from the Tikhonov reconstructions, whose free amplitude defines a learned scale
  $\gamma^\star(\lamtik)$ as the interior median $|\mvec|$.
At $\lamtikbal = 1$, the mask-confined Tikhonov reconstruction and its unmasked variant give
  $\gamma^\star\approx4.1\times10^{-4}$ and $2.5\times10^{-4}$.
All physics reconstructions use $\gamma=4\times10^{-4}$, the mask-confined value (Extended Data
  Fig.~\ref{edf:solnorm}b).

\subsection*{Synthetic validation}

\paragraph{Ground truth generation.}
The synthetic ground truth is a disk of the measured sample's dimensions (diameter
  \SI{1.1}{\micro\metre}, thickness \SI{300}{\nano\metre}) initialized from the measurement itself:
  we take the reconstruction of the full measurement without the energy term ($\lamE = 0$), a unit
  field on the $\SI{4.98}{\nano\metre}$ physics grid (Extended Data Fig.~\ref{edf:synth}a).
We relax this state with the Landau--Lifshitz--Gilbert
  equation~\cite{landau_theory_1935,gilbert_phenomenological_2004} in
  NeuralMag~\cite{abert_neuralmag_2025} to
  equilibrium at the Extended Data Table~\ref{tab:params} parameters with
  $D = \SI{0.66e-3}{\joule\per\metre\squared}$, and block-average it onto the tomography grid.
The relaxation turns the noisy seed into a smooth labyrinth of stripes that follows the arrangement
  of the measured texture (Extended Data Fig.~\ref{edf:synth}a).

\paragraph{Measurement simulation.}
We project this state with the forward model of Eq.~\eqref{eq:forward_methods}, the operator the
  reconstruction inverts, at the $29+38$ calibrated stage angles of the measurement and at the
  magnetization scale $\gamma$ of the reconstructions, so the measured noise amplitude carries over
  directly.
The noise is a stationary Gaussian field with the radially averaged power spectrum of air-only
  patches of the measured XMCD channel, scaled to the median standard deviation of the measured air
  pixels, and added with opposite sign to the two polarizations.

\paragraph{Evaluation.}
We reconstruct the synthetic projections using the same solver settings and disk mask as the
  experimental data.
Fidelity is quantified by the relative reconstruction error
  $\errm = \lVert \mstar - \mgt \rVert / \lVert \mgt \rVert$, evaluated inside the disk mask.

\subsection*{Use of large language models}
We used the large language model tools Claude (Anthropic) and Antigravity (Google) to assist with writing the analysis and
  reconstruction code, with the data analysis, and with drafting and editing the manuscript text.
The authors reviewed, tested, and revised all output and take full responsibility for the code,
  the results, and the content of this article.

\subsection*{Data and code availability}
The measured XMCD projections, the preprocessed solver inputs, the synthetic ground truth, and every
  reconstruction shown in the main text and the Extended Data will be archived at Zenodo upon
  journal publication.
The preprocessing and reconstruction code (the \texttt{xmcdrecon} package, JAX and NeuralMag), with
  the recipes and commands behind every measured-data reconstruction of this work, will be released
  on GitHub and archived at Zenodo upon journal publication.

\begin{acknowledgments}
We acknowledge Diamond Light Source for time on instrument I08-1 under proposal MG37535\nobreakdash-2.
This research was funded in whole or in part by the Austrian Science Fund (FWF) 10.55776/PIN1629824
  and 10.55776/P34671.
P.M., R.Y., M.R.F., and C.D. acknowledge funding from the Max Planck Society Lise Meitner
  Excellence Program and funding from the European Research Council (ERC) under the ERC Starting
  Grant 3DNANOQUANT 101116043.
A.F.-P. acknowledges support by the European Community under the Horizon 2020 Program, Contract
  No.~101001290 (3DNANOMAG).
The crystal growth at Warwick was supported by two Engineering and Physical Sciences Research
  Council grants: EP/Z535874/1, ``Single Crystal Growth at Warwick'', and EP/Z534067/1,
  ``Controlling the Formation and Properties of Topological Magnetic Phenomena''.
For the purpose of open access, the authors have applied a CC BY public copyright license to any
  Author Accepted Manuscript version arising from this submission.
\end{acknowledgments}

\subsection*{Author contributions}
A.S. developed the physics-informed reconstruction framework, carried out the reconstructions and
  the analysis, and wrote the manuscript.
P.M. prepared the samples, performed the X-ray ptychographic tomography measurements, the
  ptychographic reconstruction and coarse alignment of the projections, and contributed to the
  magnetic reconstructions and the analysis.
L.J.R.H. and B.K. set up and operated beamline I08-1 and supported the ptychographic tomography
  measurements.
L.T., R.Y., and M.R.F. contributed to the X-ray measurements.
D.A.M. and G.B. grew the single crystals.
A.C. contributed to the implementation of the reconstruction method.
D.S. contributed to the supervision of the project.
A.F.-P. and M.G.-S. contributed to the conception and planning of the project and to writing the
  proposal.
C.D. supervised the experimental work.
C.A. is the lead developer of the underlying NeuralMag micromagnetic framework and contributed to
  and supervised the development of the reconstruction method.
All authors discussed the reconstructions, the analysis, and the results, and commented on the
  manuscript.

\subsection*{Competing interests}
The authors declare no competing interests.

\bibliography{references}

\clearpage
\onecolumngrid
\setcounter{figure}{0}
\renewcommand{\figurename}{Extended Data Fig.}
\setcounter{table}{0}
\renewcommand{\tablename}{Extended Data Table}
\section*{Extended Data}

\begin{table}[h]
\centering
\caption{\textbf{The two datasets of this work, with their material parameters, acquisition,
  and discretization.}
The upper block lists the literature parameters of Ref.~\cite{birch_toggle-like_2022}, the DMI
  constants, and the two grids.
The lower block lists the acquisition and geometry of each dataset, including the zero-point
  offset added to the logged stage angles of each measured series (Methods, 'Data acquisition and
  preprocessing').
The synthetic dataset is described in Methods, 'Synthetic validation'.
The last two rows list the figures that use each dataset.}
\label{tab:params}
\scriptsize
\begin{tabular}{l c c}
\hline\hline
 & experimental & \swatch{specsynth}synthetic \\
\hline
Saturation magnetization $\Ms$ & \multicolumn{2}{c}{\SI{460.55}{\kilo\ampere\per\metre}} \\
Exchange stiffness $A_0$ & \multicolumn{2}{c}{\SI{5.729}{\pico\joule\per\metre}} \\
Bulk DMI constant $D_0$ (literature) & \multicolumn{2}{c}{\SI{0.60e-3}{\joule\per\metre\squared}} \\
Bulk DMI constant $D$ (readout, used in all reconstructions) & \SI{0.66e-3}{\joule\per\metre\squared} & \SI{0.66e-3}{\joule\per\metre\squared} (truth) \\
Tomography cell size & \multicolumn{2}{c}{$\approx\SI{9.96}{\nano\metre}$} \\
Physics cell size & \multicolumn{2}{c}{$\approx\SI{4.98}{\nano\metre}$ ($2\times$ refine)} \\
Stored projection window & \multicolumn{2}{c}{$250\times250$ px} \\
\hline
Thickness & \multicolumn{2}{c}{\SI{300}{\nano\metre}} \\
Diameter & \multicolumn{2}{c}{\SI{1.1}{\micro\metre}} \\
\sxz{} (projections, range) & \multicolumn{2}{c}{$29$, $[-41,+15]^\circ$} \\
\syz{} (projections, range) & \multicolumn{2}{c}{$38$, $[-30,+47]^\circ$} \\
Projections & measured & simulated at the measured angles with measured noise \\
Angle offset (\sxz, \syz) & $15^\circ$, $33^\circ$ & -- \\
Reconstruction box (tomography cells) & \multicolumn{2}{c}{$120\times120\times30$} \\
Physics grid (cells) & \multicolumn{2}{c}{$240\times240\times60$} \\
Face-on stripe period (nm) & $120$ & $122$ \\
\hline
Main figures & 1--4 & -- \\
Extended Data figures & 2--5 & 1 \\
\hline\hline
\end{tabular}
\end{table}

\begin{figure}[h]
\centering
\import{figures/}{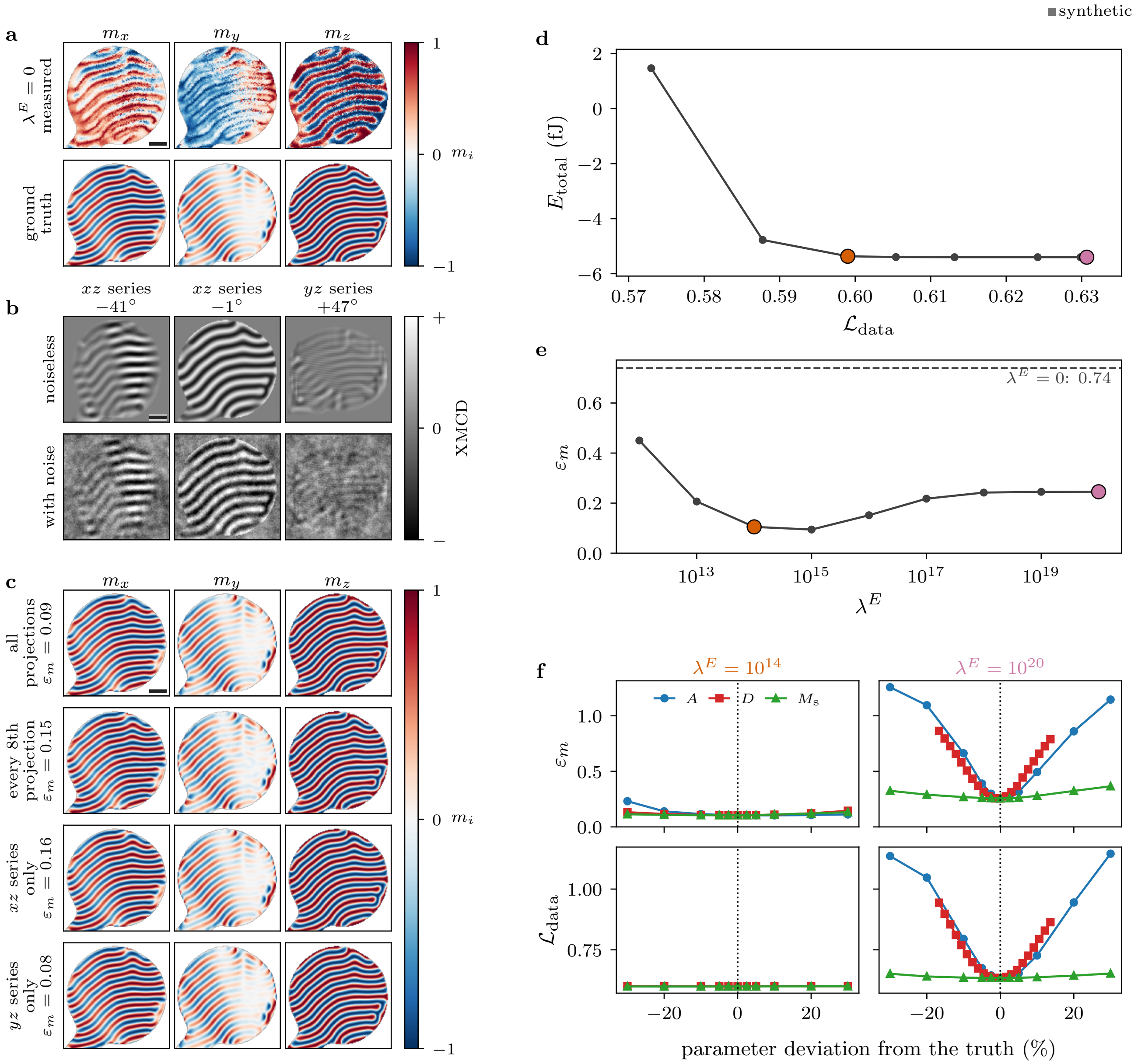}
\caption{\textbf{Synthetic validation of both operating regimes.}
Synthetic projections are generated from a known magnetization state with noise and reconstructed
  with the experimental protocol (Methods, 'Synthetic validation').
\textbf{a},~The reconstruction of the measured disk without the energy term ($\lamE = 0$, top)
  and the ground truth relaxed from it (bottom), components $\mx$, $\my$, $\mz$ in the disk plane.
\textbf{b},~Input XMCD projections, noiseless (top) and with noise (bottom), for three views: the
  \sxz\ at its largest tilt, normal incidence, and the \syz\ at its largest tilt.
\textbf{c},~The reconstructions of the four acquisitions of
  Fig.~\ref{fig:projthin} at $\lamE = 10^{15}$, laid out as Fig.~\ref{fig:projthin}b.
Each row label gives the relative error $\errm$ to the truth.
\textbf{d},~L-curve of the full acquisition, micromagnetic energy $\Etot$ versus data misfit $\Ldata$
  for $\lamE$ from $10^{12}$ to $10^{20}$, with $\lambal$ and $\lamhigh$ indicated.
\textbf{e},~$\errm$ of the full acquisition versus $\lamE$, with $\lambal$, $\lamhigh$ and the
  value at $\lamE = 0$ indicated.
\textbf{f},~Reconstructions with $A$, $D$, or $\Ms$ set to a wrong value, in the balanced regime
  ($\lambal$, left) and the readout regime ($\lamhigh = 10^{20}$, right, $D$ on the refined grid of
  Fig.~\ref{fig:adgrid}a).
The top row shows $\errm$ and the bottom row the data misfit $\Ldata$.
Scale bars, \SI{200}{\nano\metre}.}
\label{edf:synth}
\end{figure}

\begin{figure}[h]
\centering
\import{figures/}{residuals.pgf}
\caption{\textbf{Measured, modeled and residual projections.}
Measured XMCD projections $y$ (top row of each block), the projections $\pvec(\mstar)$ of the
  reconstruction at $\lambal$ (middle) and the residual $y - \pvec(\mstar)$ (bottom), for the \sxz\
  (upper block) and the \syz\ (lower block), on one gray scale.
Each block shows five single frames spread evenly over the tilt range in which the stripes are
  resolved: normal incidence, the ends of that range, and the frames halfway to either end.
Numbers in the residual panels give the frame-wise relative residual $\lVert \pvec_i(\mstar) - \yvec_i \rVert / \lVert \yvec_i \rVert$, whose root-mean-square over all $N$ frames is $\Ldata = 0.83$ (Eq.~\eqref{eq:ldata}).
Scale bar, \SI{200}{\nano\metre}.}
\label{edf:residuals}
\end{figure}

\begin{figure}[h]
\centering
\begin{subfigure}[t]{0.49\textwidth}
\panellabel{a}
\centering
\import{figures/}{dselect_lambda_saturation.pgf}
\end{subfigure}\hfill
\begin{subfigure}[t]{0.49\textwidth}
\panellabel{b}
\centering
\import{figures/}{dselect_gamma_robust.pgf}
\end{subfigure}
\caption{\textbf{Sensitivity of the $D$ selection to the energy weight $\lamE$ and the
  magnetization scale $\gamma$.}
Data misfit $\Ldata$ versus $D$, as in the $D$ selection of Fig.~\ref{fig:adgrid}.
\textbf{a},~Curves for a sweep of $\lamE$ at the default $\gamma$, colored by $\log_{10}\lamE$.
The dotted line marks the literature $D_0$.
At small $\lamE$ the data term predominately shapes $\mvec$ and no dip forms.
The dip sharpens through the intermediate regime and freezes past the knee
  ($\lamE \gtrsim 10^{17}$), where $\mvec$ settles into the energy minimum nearest to the data and the curves for
  the highest decades coincide.
The selection weight $\lamhigh = 10^{20}$ lies in this regime.
There the minimum stays at $D = \SI{0.66e-3}{\joule\per\metre\squared}$ on the
  $\SI{0.03e-3}{\joule\per\metre\squared}$ grid of this sweep, the value read off the refined
  grid of Fig.~\ref{fig:adgrid}a.
\textbf{b},~The same curves at $\lamhigh$ for $\gamma$ between \num{1e-4} and \num{6e-4}, with
  $D_0$ dotted as in \textbf{a}.
A larger $\gamma$ deepens the dip and a smaller $\gamma$ flattens it.
The minimum stays at $D = \SI{0.66e-3}{\joule\per\metre\squared}$ for every $\gamma$.}
\label{edf:dsens}
\end{figure}

\begin{figure}[h]
\centering
\import{figures/}{solnorm.pgf}
\caption{\textbf{Tikhonov reconstructions and the magnetization scale $\gamma$.}
\textbf{a},~L-curve of the Tikhonov sweep, solution norm $\lVert\mvec\rVert$ versus the data
  misfit $\Ldata$, for the unmasked field on the full box and for the mask-confined field, with
  $\lamtikbal = 1$ marked.
\textbf{b},~The learned magnetization scale $\gamma^\star$ (interior median $|\mvec|$) versus
  $\lamtik$ for the same two sweeps, $\lamtikbal$ marked as in \textbf{a}.
$\gamma^\star$ changes monotonically with $\lamtik$.
The fixed $\gamma$ of the physics reconstructions (dashed line, Methods, 'Magnetic reconstruction
  framework') is the mask-confined value at $\lamtikbal$.
\textbf{c},~The unmasked reconstruction at $\lamtikbal$, components $\mx, \my, \mz$ in the disk plane
  ($xy$, left) and a side view ($yz$, right), shown in full with the mask boundary drawn gray dashed.
The field smears across the full depth and leaks outside the mask.
\textbf{d},~The mask-confined reconstruction at $\lamtikbal$, in the same layout.
The masked volume carries the same in-plane stripe pattern at higher contrast, and the amplitude
  rises from $\gamma^\star = 2.5\times10^{-4}$ (unmasked) to $4.1\times10^{-4}$.
Each reconstruction is shown on one color scale, set by its $99.5$th percentile over all components.
Scale bars, \SI{200}{\nano\metre}.}
\label{edf:solnorm}
\end{figure}

\begin{figure}[h]
\centering
\begin{subfigure}[t]{0.42\textwidth}
\panellabel{a}
\centering
\resizebox{\textwidth}{!}{\begin{tikzpicture}
\begin{axis}[name=ax, xlabel={$\Ldataabs$}, ylabel={$\Ltv$}, width=\textwidth, tick label style={font=\footnotesize}, label style={font=\small}, scaled x ticks=false, ymin=0, xticklabel style={/pgf/number format/.cd,fixed,fixed zerofill,precision=2}]
\addplot[black, mark=*, mark size=1.3pt] coordinates {(6.621179e-02,5.594273e-03) (6.709218e-02,1.908954e-03) (6.884115e-02,9.681612e-04) (7.474725e-02,7.366124e-04) (1.242569e-01,4.987473e-04)};
\addplot[black, only marks, mark=*, mark size=2.6pt] coordinates {(6.621179e-02,5.594273e-03)};
\node[black, anchor=west, font=\small] at (axis cs:6.621179e-02,5.594273e-03) {$\lamTV{=}0.01$};
\addplot[red, only marks, mark=*, mark size=2.6pt] coordinates {(6.884115e-02,9.681612e-04)};
\node[red, anchor=south west, font=\small] at (axis cs:6.884115e-02,9.681612e-04) {$\lamTV{=}1$};
\addplot[black, only marks, mark=*, mark size=2.6pt] coordinates {(1.242569e-01,4.987473e-04)};
\node[black, anchor=south east, font=\small] at (axis cs:1.242569e-01,4.987473e-04) {$\lamTV{=}100$};
\end{axis}
\end{tikzpicture}}\\[4pt]
\panellabel{b}
\import{figures/}{charge_slices3d.pgf}
\end{subfigure}\hfill
\begin{subfigure}[t]{0.56\textwidth}
\panellabel{c}
\centering
\import{figures/}{charge_mask_overlay.pgf}
\end{subfigure}
\caption{\textbf{Validation of the disk mask through total-variation (TV) tomography of the
  electronic channel.}
\textbf{a},~L-curve: total variation $\Ltv$ versus the electronic-channel data misfit $\Ldataabs$ as
  the TV weight $\lamTV$ is swept.
The misfit is flat up to the operating point $\lamTV{=}1$ (red) and rises beyond it, where
  the reconstruction over-smooths.
\textbf{b},~The disk mask used in all magnetic reconstructions, shown in three dimensions with the three slice planes of \textbf{c}: the disk
  plane ($xy$) at mid-slab and the two side views ($xz$, $yz$) through the disk center.
\textbf{c},~Reconstructed density at $\lamTV{=}0.01$, $1$ (operating point), and $100$, shown in
  the disk plane ($xy$, top row) and the two side views ($xz$, $yz$).
At $\lamTV{=}0.01$ the density is streaky, and at $\lamTV{=}100$ it is over-smoothed.
The density extends beyond the slab in $z$ because of the missing wedge, more so at larger
  $\lamTV$.
Scale bar, \SI{200}{\nano\metre}.}
\label{edf:charge}
\end{figure}

\end{document}